# THE PHYSICAL ORIGIN OF MASS AND CHARGE II

Erik A. Haeffner
Eskadervägen 48, 183 58 TÄBY, SWEDEN, haeffner@algonet.se

This article is an alternative version of a document which already can be found on the net. (http://www.algonet.se/~haeffner). The scientific message is, however, now presented on a somewhat lower abstraction level, with the ambition that wellknown and generally accepted experimental facts may be identified more convenient and obvious as direct consequences of a new physical concept here disclosed and discussed.

The new concept CER (Condensed Electromagnetic Radiation), proposed in this article, indicates an electromagnetic origin of mass particles, in fact, an overwhelming amount of experimental evidence confirms that the CER concept is fundamental for the physical explanation of mass particle properties.

Two well established physical phenomena may, among others indicated in the manuscript, be pointed out as conclusive for the new theory:

–Parametric down conversion is a commonly used expression to describe the phenomenon when a light wave passing an optically active crystal is converted into two new oppositely polarized light waves. According to the new theory it should in consequence be possible to produce two new CER particles if the initial electromagnetic radiation, EMR, has a high enough energy (frequency) equivalent with the sum of the energies (masses) of the new particles. The pair production phenomenon (electron and positron) is thus explained by the CER concept.

–The second apparent proof is the existence of the first generation leptons and quarks. These may be predicted as CER particles, each produced by the superposition of two plane polarized EMR at right angle with a quantum phase difference of $\pi/6$ as described and illustrated in the manuscript. A consequence of the new physical interpretation of mass is the conclusion that all material in the universe is of electromagnetic origin, either in the form of EMR : $E = h\upsilon$  or as CER : $E = h\upsilon = mc^2$  (mass particles)

**Introduction**

The concept of mass has a central and fundamental position in the science of physics. Quantities such as charge and energy are correlated to mass. Electrical charge does not exist without mass, in fact, it might be called a property of mass which in itself is a form of energy, as discovered by Einstein.

The three most important theories of modern physics; Quantum Mechanics, Special Relativity and the Standard Model cannot, however, give a straightforward explanation of the physical nature or the origin of mass.

In **Quantum Mechanics** mass is described both in the form of wave equations and as point particles where the energy content is concentrated to a mathematical point. The waves are then interpreted to express the probability to find a particle within a certain space-time region. As pointed out in a study by Marmet ( 2 ) it is not easy to find a logical definition of Quantum Mechanics as a theory of physics and in this rather diffuse situation to determine if the concept of mass has a physical meaning at all. The negative opinion of Einstein in considering the physical reality of Quantum Mechanics is wellknown but to a large extent ignored. He has, however, given a good **reason** for his opinion. "I am, in fact, firmly convinced that the essentially statistical character of contemporary quantum theory is solely to be ascribed to the fact that this theory operates with an incomplete description of physical systems " ( 3 )

There has been many other efforts in the history of physics to introduce mathematical models of wave structures intended to decribe mass particles. Among those with a wide scientific background is a paper by M. Wolff. (6) Difficulties arise often, however, when the mathematical models are compared with experimental facts or when the nature and origin of waves should be explained.

The **Standard Model** is a classification system in which all elementary particles are ordered in families, generations and classes according to which forces keep particles together. The internal physical nature and structure of most fundamental particles - leptons and quarks - is so far not known and not needed for the classification as such but of course much wanted for the understanding of nature.

**Special Relativity** is a rational and logical description of particle physics. Masses exist independently of an observer. Relativistic corrections are used to give a description independent of the observers velocity. This rational mathematical description of nature conforms with experimental facts but cannot provide a physical explanation of the existence of mass particles.

**Summarizing:** None of these mentioned three theories can, severally or in combination, explain the nature or the origin of mass. For that purpose an additional physical theory is consequently needed.

In this article a new physical concept Condensed Electromagnetic Radiation, CER, is introduced with the aim to fill this gap and to constitute a physical explanation of the nature and the origin of mass and charge. The CER concept will be compared with physical realities, such as the existence of fundamental mass particles and scientifically accepted experimental facts.

**The CER Concept**
Following wellknown physics we find that in a linear or plane polarized electromagnetic radiation, EMR, the electric and magnetic field strength vectors oscillate in planes perpendicular to each other and to the direction of propagation. If we superimpose two independent electrical oscillations we can produce in different ways polarized EMR. If the two oscillations are in phase and at right angle we find as a result a new plane polarized electromagnetic radiation. Should the two initial EMR:s still have the same amplitude but there exists a phase difference of 90 degrees, circular polarized electromagnetic radiations will appear. With a phase difference of 30 degrees or 150 degrees the radiation is said to be elliptically polarized with the electric vectors rotating respectively to the right and to the left when looked upon in the direction of radiation propagation.( fig 1., fig 2.)

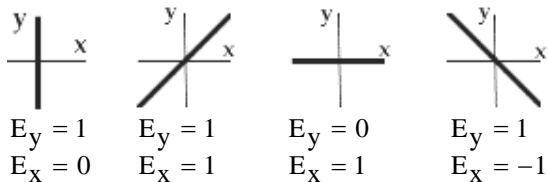

$E_y = 1$    $E_y = 1$    $E_y = 0$    $E_y = 1$
$E_x = 0$    $E_x = 1$    $E_x = 1$    $E_x = -1$

Fig 1. Superposition of X-vibrations and Y-vibrations in phase.

If the initial vibrations are:
$E_y = A \sin(\omega\tau + \phi)$, where
$\omega$ = frequency in radians/s
$\tau$ = time in seconds
$\phi$ = phase difference, and
$E_x = B \sin(\omega\tau)$ ; assuming
A = B = the amplitude of vibrations, we find figures of polarization as shown on fig 2.

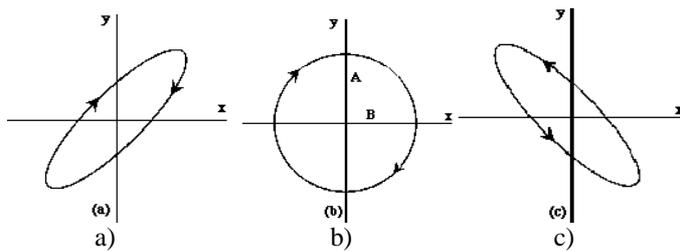

a)      b)      c)

Fig 2. Superposition at right angle of oscillations Ex and Ey with the same amplitude at phase differences:
a) $\phi = \dfrac{\pi}{6}$    b) $\phi = \dfrac{\pi}{2}$    c) $\phi = \dfrac{5\pi}{6}$

**The endpoints of electrical vectors** in a circularly polarized EMR moving in space can be illustrated by a spiral spring, fig 3.

Circular Polarization

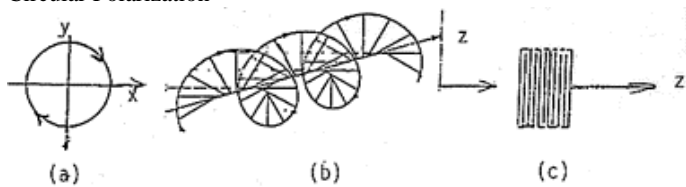

Elliptical Polarization

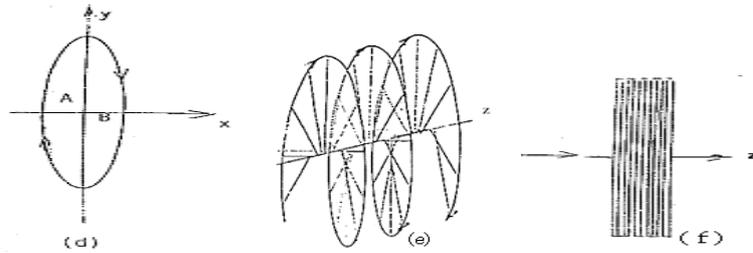

Fig 3 Formation (a,d), propagation (b,e) and condensation (c,f) of circulary and elliptically polarized wave packets, CER.

If the frequency is high the distance between subsequent threads of the spring is small. We now foresee –**the hypothesis**– that at a certain small distance the individual electromagnetic wave units are squeezed into a closely packed wave packet, which we call a **Condensed Electromagnetic Radiation,** CER, and identify as a CER mass particle. In the same way we can produce elliptically polarized CER mass particles. (It should be noted that the expression "condensed" here means a condensation in space-time, not to be confused with ordinary gas condensation. It should also be noted that we here interpret Planck´s constant **h** as the energy content of one EMR wave unit. If we multiply **h** with the frequency $\upsilon$, we will get an amount of energy $h\upsilon$, which has been called a **photon** or elderly a **"light quanta"** (energy quantum). These EMR energy quantities are not massparticles. The CER concept, on the other hand, describes how EMR wave units after circular or elliptical polarization are transformed into mass particles. We can calculate the number of wave units in an electron by dividing the rest energy of the electron with Planck´s constant, and find that the electron as a CER particle contains $1.2356 \times 10^{20}$ wave units. )

The CER concept of circularly or elliptically polarized electromagnetic radiation in a condensed state as a physical identification of mass, is discussed in the following sections and investigated as how this concept agrees with wellknown and scientifically accepted physical phenomena.

**Leptons and quarks as CER particles**
Through the theoretical and experimental development of the standard model, the plurality of elementary particles have slowly found their place in a framework of families, classes and generations. Leptons and quarks of the first generation, fig 4, are now assumed to be the real fundamental particles from which all the other "elementary" particles originate either as excitation states or as composite mass particles.
It is then obviously important to investigate how the CER hypothesis will conform with the symmetry and properties of these mentioned most fundamental particles. Fig 4 illustrates polarized EMR waves, each composed of two linear polarized EMR's with phase differences in 4 steps, $\frac{3\pi}{6};\frac{2\pi}{6};\frac{\pi}{6};0$ quadrant I, followed by the 4 symmetrically opposite phase difference angles in quadrant III. We might describe the phase difference going from 0°, in steps of $\frac{\pi}{6}(30°)$ to $\frac{3\pi}{6}(90°)$ and from $\frac{6\pi}{6}(180°)$ to $\frac{9\pi}{6}(270°)$ as increasing the degree of polarization of the resultant EMR.

In the CER concept it is assumed that at a certain small distance between adjacent electrical vector spiral units in circular and elliptical EMR (fig 3) these superimpose to compress the EMR into CER mass particles. It would thus be possible to identify the existing most fundamental particles with the illustrated species of polarized CER. The two bottom lines of fig 4 indicate the first generation of leptons and quarks with their electrical charges. We find that these most fundamental particles of the standard model may easily be identified as CER particles by their symmetry and charge which increases in polarization steps of $\frac{\pi}{6}(30°)$

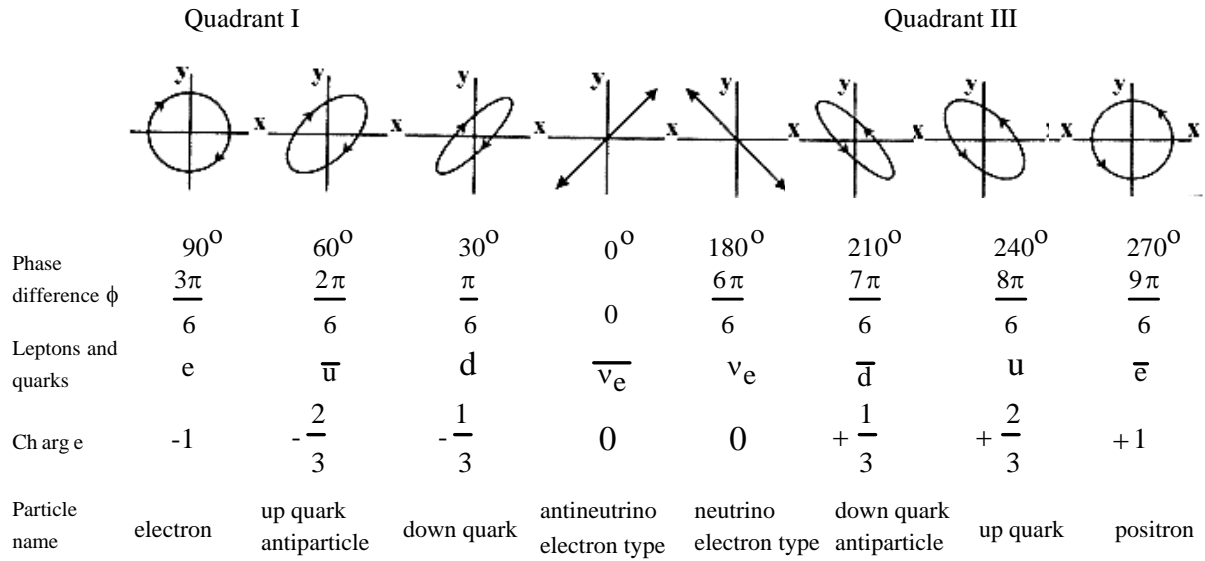

| | Quadrant I | | | | Quadrant III | | | |
|---|---|---|---|---|---|---|---|---|
| Phase difference φ | $90^o$ $\frac{3\pi}{6}$ | $60^o$ $\frac{2\pi}{6}$ | $30^o$ $\frac{\pi}{6}$ | $0^o$ $0$ | $180^o$ $\frac{6\pi}{6}$ | $210^o$ $\frac{7\pi}{6}$ | $240^o$ $\frac{8\pi}{6}$ | $270^o$ $\frac{9\pi}{6}$ |
| Leptons and quarks | e | $\overline{u}$ | d | $\overline{\nu_e}$ | $\nu_e$ | $\overline{d}$ | u | $\overline{e}$ |
| Charge | -1 | $-\frac{2}{3}$ | $-\frac{1}{3}$ | 0 | 0 | $+\frac{1}{3}$ | $+\frac{2}{3}$ | +1 |
| Particle name | electron | up quark antiparticle | down quark | antineutrino electron type | neutrino electron type | down quark antiparticle | up quark | positron |

Fig 4 Superposition at right angle of two plane polarized EMR waves with the same amplitude but with phase difference of $n\frac{\pi}{6}$ where n = 0,1,2,3 (quadrant I) and 6,7,8,9 (qadrant III).

**The resultant CER wave packets are identified in symmetry and in electric charge with leptons and quarks of the first generation.**

This identification will also explain the charge unit of the electron and the positron as corresponding to the maximum polarization possible. The quarks appear as elliptically polarized CER wave packets with different polarization degrees. The electron neutrinos are each formed of two superimposed plane polarized EMR at right angle and in phase. They consequently have a mass according to the fundamental equation

$$h\nu = mc^2$$
$$m = \frac{h\nu}{c^2}$$

The explanation of the old problem why the electron has the experimentally well determined charge **e**, and nothing else, higher or lower, is then simply that it has the maximum polarization degree possible as a CER particle.

## *Confirmation of the new theory by experimental facts*:
**Pair Production - Annihilation**
If an EMR travels into a crystal that has nonlinear optical properties, two new EMR:s with opposite angular momentum are created, that is one clockwise and one anticlockwise circularly polarized electromagnetic radiations. The sum of the energies of the two new radiations equals exactly the energy of the parent EMR. The phenomenon is called spontaneous parametric down conversion.
According to the CER concept as described above the secondary polarized radiations should at a certain
frequency be converted to CER mass particles. That is also the case in physical reality, when a high energy EMR > 1.02 MeV, in the vicinity of a nucleus is transformed into one electron (mass energy 0.51 MeV) and its antiparticle the positron with the same mass. The reaction is wellknown as pair production and may physically be interpreted as a spontaneous down conversion of a high energy electromagnetic radiation. **The pair production phenomenon is an experimental confirmation of how CER mass particles are produced by the condensation of high energy circular polarized electromagnetic radiation.**

**Relativistic mass effects**
"A real world", or in other words, a physical realism implies or has a necessary prerequisite that mass and electromagnetic radiation exist independently of an observer. Einsteins discovery or deduction that the velocity of EMR is a constant **c** in vacuum and independent of any observers own velocity is, in fact, the physical foundation of all relativistic phenomena.
A CER mass particle is physically a compressed or condensed wave packet of circularly or elliptically polarized electromagnetic radiation which in itself has the velocity **c**. If we assume that the wave packet is moving with a velocity **v** relative to the constant velocity **c** we find that when calculating the frequency ω (radians per s) of the polarized electromagnetic radiation constituting the CER wave packet we must consider the Doppler effect in order to determine the shift of frequency when the wave packet with a velocity **v** is moving relative to **c**. When the velocity **v** is increasing the frequency ω is also increasing according to the formula:

$$\omega_1 = \frac{\omega_0 \left(1 + \frac{v}{c}\right)}{\sqrt{1 - \frac{v^2}{c^2}}}$$

As the mass is directly proportional to the frequency according to

$$\eta \omega_0 = m_0 c^2 \quad \text{where} \quad \eta = \frac{h}{2\pi}$$

we find that the mass of a CER particle follows the formula for relativistic increase with velocity

$$m_1 = \frac{m_0 \left(1 + \frac{v}{c}\right)}{\sqrt{1 - \frac{v^2}{c^2}}}$$

The increase of mass when a particle is moving relative to the fixed EMR velocity **c** may then be interpreted as a Doppler effect increasing the frequency of the EMR which constitutes the CER wave packet. No observer is needed for this deduction as only internal physical relations are used in the **c** frame of a reference.

**Waves associated to mass particles**
A wellknown mathematical formula used to indicate wave properties of mass particles is the de Brolie wave length

$$\lambda = \frac{h}{m \cdot v}$$

h = Planck´s constant,
m = particle mass and
v = particle velocity as measured by an external observer.

This equation is physically supported by diffraction experiments where a stream of electrons pass through slites in a screen and then produce a diffration pattern on a parallell target screen. There is so far no physical explanation of the origin or nature of these mass particle waves producing the diffraction patterns. Textbook formulations are that the de Broglie waves somehow are "associated " with mass. It is also often pointed out that matter waves are not of electromagnetic origin. However, by introducing the new concept CER, Condensed Electromagnetic Radiation, we have established that mass particles really are of electromagnetic origin. **We can therefore look on the de Broglie phenomenon with diffraction patterns created by mass particles in the light of the CER concept.**

Starting with the fundamental equation $h\upsilon = mc^2$
where
h = Planck´s constant
υ = frequency of electromagnetic radiation, EMR
m = the mass of a CER wave packet
c = EMR velocity in vacuum

If we exchange υ with $c/\lambda$, where λ = wave length, we get $\lambda = \dfrac{h}{m \cdot c}$

It has been shown that the electron and the other leptons and quarks of the first generation confirm with respect to charge and symmetri their identity as CER particles. We can then treat the electron as a CER mass particle with the experimentally determined mass $m_e$ and use the formula above to calculate the wavelength $\lambda_e$ will decrease, also found by diffraction experiments.

It has been shown that the electron and the other leptons and quarks of the first generation confirm with respect to charge and symmetry their identity as CER particles. We can then treat the electron as a CER mass particle with the experimentally determined mass $m_e$ and use the formula above to calculate the wavelength $\lambda_e$ of an electron at rest. Using the known constants **h** and **c** we find $\lambda_e = 2.42 \times 10^{-10}$ cm

What happens if the electron moves towards, as an example, a diffraction chamber with a velocity $v$? A velocity of a CER particle is always a movement relative to the constant velocity **c**. As has already been shown the electron mass will then increase according to the relativity equation

$$m_1 = \frac{m_0 \left(1 + \frac{v}{c}\right)}{\sqrt{1 - \frac{v^2}{c^2}}}$$

From the equation $\lambda = \frac{h}{m \cdot c}$ we realize that when the electron mass increases its wavelength $\lambda_e$ will decrease, as also found by diffraction experiments

The CER concept can consequently provide a physical explanation of what has been called " matter waves" or "associated waves". It should be noted that there is no reference to an "observer" or the relative velocity to such an "observer" when calculating the wavelength of a mass particle.

**An alternative way to estimate the wavelength of the electron is by using the experimentally found Compton scattering relation, which describes the relation of wave lengths when an incoming EMR $(\lambda_0)$ is scattered by an electron at an angle $\theta$ from its initial direction. The scattered EMR will then get the new wavelenght $\lambda_s$ and the scattered electron is found to have the dimension of a wavelenght $\lambda_e$ according to the equation $\lambda_s - \lambda_0 = \lambda_e (1 - \cos \theta)$**

**When $\theta = 90°$ it is found that the "Compton wavelenght" $\lambda_e = \frac{h}{m_e \cdot c}$**

**which is the same value as already found if we concider the electron as a CER mass particle. This means that we have found another experimental confirmation of the CER concept.**

**Uncertainty relations**
The basic postulate of Quantum Mechanics is the Heisenberg uncertainty principle. If a complete mathematical description of fundamental physics can be built on this postulate is a question that has been discussed both by physicists and philosophers for decades. Steven Weinberg ( 4 ) and many others confirm that Quantum Mechanics provides a precise framework for calculating energies, trasition rates and probabilities. Marmet ( 2 ) and others have, on the other hand, found many absurdities when comparing the QM mathematical descriptions with a physical reality. These scientific controversies have apparently their root in a problem that already Gödel ( 5 ) discussed. The mathematics of a system ensures that there are no internal mathematical contradictions. Mathematics cannot, however deduce results from relations that are external to the mathematical system.

With the introduction of the CER concept we can end this controversy and give the Heisenberg deduction of the uncertainty principles a physical meaning. The starting point is the same relationship $E = h\upsilon$ as Heisenberg used for the mathematical formation of a Fourier wave packet. The difference is that according to the CER concept the wave packet is identified physically as compressed electromagnetic radiation. In that way we can give the foundation of Quantum Mechanics a physical reality. This is shown (ref.1) by a simple deduction of the Heisenberg uncertainty relations using the CER wave packet concept .

The question, however, to what extent further quantum theory mathematical transformations correspond to physical realities will not be discussed here.

**Conclusions**
The new concept of condensed electromagnetic radiation CER indicates a physical structure of mass particles. It has been demonstrated that the fundamental particles leptons and quarks of the first generation, which in the Standard Model are thought as building blocks of composite particles, coincide in charge and symmetry with a series of CER wave packets where the phase difference between originally plane polarized EMR waves is integer multiples of $\frac{\pi}{6}$ radians.

Besides this successful demonstration of the physical origin of mass and charge the CER concept is supported by the explanation it provides of, so far mysterious, phenomena of nature such as mass particle production, mass increase with velocity, elementary particle symmetries, charge quantization, wave structure of matter, etc. The CER concept is consequently of interest for further physical studies and experiments. What comes to mind, as a high priority, is the synthesizing of existing or new particles through the building up, or expressed in physical terminology, the superposition of high energy EMR:s which might differ in phase, frequency and amplitude. A consequence of the new physical interpretation of mass is the conclusion that all materia in the universe is of electromagnetic origin, either in the form of EMR : $E = h\upsilon$ or as CER : $E = h\upsilon = mc^2$

**The Planck - Einstein energy formulas above appears then as very fundamental equations of the universe.**

**References**

[ 1 ]  E.A. Haeffner, The Physical Origin of Mass and Charge.
http://www.algonet.se/~haeffner, (Febr. 1, 1997)
(Officially registered June 2, 1995)

[ 2 ]  P. Marmet, Absurdities in Modern Physics. A Solution. (1993), Ed. Les Editions du Nordir, c/o
R. Yergeau, 165 Waller Simard Hall Ottawa, On, Canada K1N6N5.
http://www.newtonphysics.on.ca/Heisenberg/contents.html (1997)

[ 3 ]  P.A. Schilp, ed. Albert Einstein: Philosopher - Scientist, pp 666,672
Evanston,IL: Library of Living Philosophers, (1949)

[ 4 ]  S. Weinberg, Dreams of a Final Theory.
Hutchinson Radius, London, (1993)

[ 5 ]  S.C. Kleene, The Work of Kurt Gödel. The Journal of Symbolic Logic, 41 No 4, (1976)

[ 6 ]  M. Wolff, "Beyond the Point Particle - A Wave Structure for the Electron",
Galilean Electrodynamics 6 (5), 83-91 (1995)

Acknowledgement.

The author sincerely thanks the Editor Dr Cynthia K. Whitney for her kind
advice and support during the publishing procedure.